\documentclass{aa}
\usepackage{graphicx}
\usepackage{txfonts}
\begin{document}
   \title{Nuclear star formation in the quasar PG1126$-$041 from adaptive
   optics assisted spectroscopy \thanks{Based on data obtained
   at the VLT through the ESO program 71.B-0453(A).}}

   \author{G. Cresci \inst{1} \and R. Maiolino \inst{2} \and A. Marconi \inst{2}  
          \and F. Mannucci \inst{3} \and G.~L. Granato \inst{4}}

   \offprints{G. Cresci}

   \institute{Dip. di Astronomia - Universit\`a di Firenze,
   	Largo E. Fermi 5, I-50125 Firenze, Italy;
              \email{gcresci@arcetri.astro.it}
         \and INAF, Osservatorio Astrofisico di Arcetri, 
           Largo E. Fermi 5, I-50125, Firenze, Italy
	 \and Istituto di Radioastronomia, sezione di Firenze,
           Largo E. Fermi 5, I-50125, Firenze, Italy
	 \and INAF, Osservatorio Astronomico di Padova, 
	 Vicolo dell'Osservatorio 5, I-35100 Padova, Italy}

   \date{Received ; accepted }

   \abstract{We present adaptive optics assisted spectroscopy of three quasars
   obtained with NACO at VLT. The high angular resolution achieved with the
   adaptive optics ($\sim0.08''$), joined to the diagnostic power of near-IR
   spectroscopy,
   allow us to investigate the properties of the innermost 100~pc of
   these quasars.
   In the quasar with the best adaptive optics correction, PG1126$-$041, 
   we spatially resolve the Pa$\alpha$ emission within the
   nuclear 100~pc. The comparison with higher excitation lines suggests that
   the narrow Pa$\alpha$ emission is due to nuclear star formation.
   The inferred intensity of the nuclear star formation
   (13~$\rm M_{\odot}~yr^{-1}$) 
   may account for most of the far-IR luminosity observed in this quasar.

   }

   \maketitle
%

\section{Introduction}

The study of the stellar populations and star formation
in the host galaxies of Active Galactic Nuclei (AGNs) is 
fundamental to understand the connection between black-hole
growth and galaxy formation.
Several evidences for such a connection have been obtained in
local low-luminosity AGNs, i.e. Seyfert galaxies
(e.g. Cid Fernandes et al. \cite{cid01}, Heckman et al. \cite{heckman97},
Maiolino et al. \cite{maiolino95}, Oliva et al. \cite{oliva99}).
At higher, quasar-like luminosities the
investigation of star formation and stellar population is much
more difficult because the tracers of star formation are heavily
diluted by the strong active nucleus. Nevertheless a few observational programs
have been successful in disentangling the host galaxy
and star formation from the quasar light. The separation
of the two components has been achieved either through a
spectroscopic decomposition or through high angular resolution
imaging observations (e.g.
Canalizo \& Stockton \cite{canalizo01},
Courbin et al. \cite{courbin02}, Dunlop et al. \cite{dunlop03},
Jahnke et al. \cite{jahnke03}, Schade et al. \cite{schade00}).
By using such observating strategies,
evidence both for active star formation and relatively
quiescent hosts was found. However, all these studies probed 
the stellar population on scales of a few kpc, while the stellar
activity within the central 100~pc, where any interplay with the quasar
activity is expected to occur, remains poorly explored.

During recent years the performances of adaptive optics systems have improved
significantly,
achieving very high angular resolutions ($\la 0.1''$) even on relatively 
faint sources. 
Such high angular resolution, coupled with spectroscopic capabilities
of some of these systems, allows us to efficiently disentangle components
associated with stellar activity even in the circumnuclear region of
luminous quasars. In this letter we present NACO observations of
three quasars at z $\sim$0.06 selected to have an excess of far-Infrared
emission. In one of the quasars we find evidence for star formation
within the central 100 pc with a rate of $\rm \sim 13~M_{\odot}~yr^{-1}$, 
which may account for most of its far-IR emission.
We assume a "standard" cosmology, with $\rm H_0 = 70~km~s^{-1}~Mpc^{-1}$,
$\rm \Omega _m =0.3$ and $\rm \Omega _{\Lambda}=0.7$.


\section{Sample selection, observations and data reduction}

The targets were selected from the sample
of Andreani et al. (\cite{andreani01}) who provide mid-IR and far-IR data for a large number
of optically selected quasars.
We selected quasars matching the following constraints:
1) z$\sim$0.06, so that the CONICA projected resolution
on the source is better than 100~pc (in the K band) and, at the
same time, both Pa$\alpha$ and [SiVI]1.97$\mu$m emission lines
are shifted into the K band; 
2) far-IR luminosity $\rm L_{FIR}>10^{44}~erg~s^{-1}$, i.e.
objects more IR luminous than the average of quasars at the same
redshift and, therefore, suspected to harbor starburst activity;
3) radio quiet, to avoid synchrotron contamination to the far-IR
radiation; 
4) visible magnitude V$<$14.5 mag, so that the Adaptive Optics correction can
achieve good performances with Strehl ratios of $\sim$0.2 
by using the quasar itself as a reference to close the loop.
Four quasars were selected and three of them, whose
properties are listed in Tab.~1, were observed in service mode with NACO.
\begin{table*}
	\label{osservazioni}
	\begin{center}
	\begin{tabular}{l c c c c c c c c c}
	\hline
	\hline
	Name & RA(J2000) & DEC(J2000) & z & m$_V$ & L$_{\rm FIR}$ & Date & seeing & FWHM & SR \\
	\hline
	IRAS09149$-$6206 & 09 16 09.4 & $-$62 19 29 & 0.057 & 14.1 & 7.9 & Apr 20, 2003 & 1.0$''$ & 0.09$''$ (100 pc) & 5\% \\
	PG1126$-$041 & 11 29 16.6 & $-$04 24 08 & 0.060 & 14.5 & 3.2 & Jun 17, 2003 & 0.4$''$ & 0.07$''$ (81 pc) & 26\% \\
	PG2214+139 & 22 17 12.3 & +14 14 21 & 0.066 & 14.5 &  1.5 & Jun 18, 2003 & 0.8$''$ & 0.09$''$ (114 pc) & 11\% \\
	\hline
	\hline
	\end{tabular}
	\caption{List of the bserved quasars, along with their properties and log of
	the observations. L$_{\rm FIR}$ is in units of
	$\rm 10^{44}erg~s^{-1}$;
	\textit{seeing} is the median DIMM seeing 
	during the observation;
	\textit{FWHM} is the Full Width Half Maximum of the PSF on the Adaptive Optics corrected images, in arcsec 
	and in parsec; \textit{SR} is the Strehl Ratio measured on the images.}
	\end{center}
\end{table*}
Both images and spectra were obtained in the K band. Images were
obtained in the IB\_2.18 filter (to avoid saturation)
with the S27 camera, yielding a pixel scale of 
$0.027''$/pix. Spectra were obtained with a $0.086''$ slit
at PA$=$0$^{\circ}$, with the Grism2 
coupled with the SK order sorting filter,
yielding a spectral resolution $\rm R=\lambda/\Delta \lambda=1400$,
and with the S54 camera yielding a pixel scale along the slit of $0.054''$/pix.
During each observing run we obtained a diffraction limited nuclear
spike  of $\sim0.08''$ (corresponding to $\sim$100~pc projected on these
quasars). In the spectroscopic mode the pixel size
undersamples the PSF. The Strehl Ratio ranges from 5\% to 26\% (see Tab.~1).

The spectra were reduced following the standard steps for NIR spectroscopy. The spectra obtained 
at different positions along the slit were subtracted from each other to remove the background, 
then flat fielded, aligned, co-added and calibrated.  
Atmospheric features were corrected by dividing for the spectrum of a reference star. 
B3III and B5V stars were used for PG1126 and IRAS09149 respectively, and
in these cases
the spectra were then multiplied by a blackbody at 20000~K to
re-establish the correct slope of the 
continuum. For PG2214 we used a G0V star, and the spectra were then
multiplied by the 
solar spectrum to remove the stellar features and continuum slope
(see Maiolino et al. \cite{maiolino96} for details). Images were instead
reduced using the \textit{eclipse} software and the CONICA pipeline.

\section{Analysis and results}

\begin{figure}[!h]
	\centering
	\resizebox{\hsize}{!}{\includegraphics{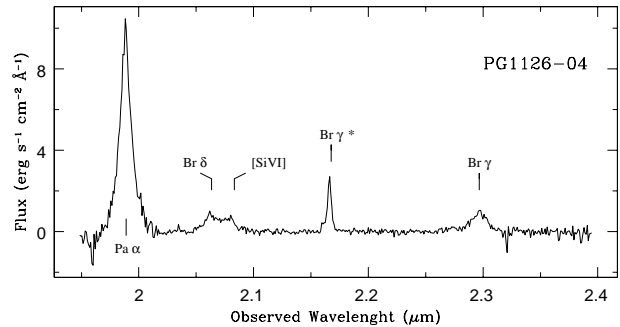}}
	\caption{Continuum subtracted spectra of PG1126$-$04. The Br$\gamma$* 
	introduced by the division of the reference B star is marked.}
        \label{spectra}
\end{figure}

In this letter we focus on the analysis of the spectra, since they provide the most
interesting information.
Of the three quasar spectra we obtain spatially resolved features only for the
case 
with the best Adaptive Optics correction, i.e. PG1126$-$041 which was observed with a
Strehl Ratio of 26\%. The integrated spectrum is shown in Fig.~\ref{spectra}. In the 
following we will concentrate on the analysis of this spectrum. 

\begin{figure}
	\centering
	\resizebox{\hsize}{!}{\includegraphics{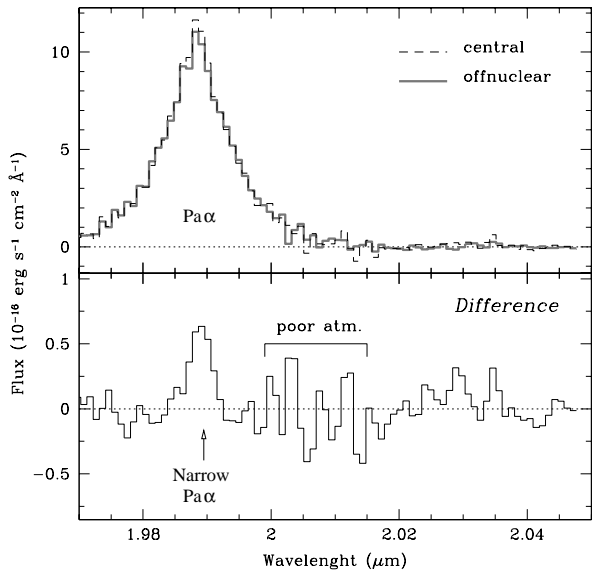}}
	\caption{Comparison between the central spectrum and the spectrum $0.054''$
	to the north, around 
	the Pa$\alpha$ region.
	In the \textit{upper panel} the solid gray line shows the (continuum subtracted) 
	central spectrum, while the black dashed line shows the rescaled off-nuclear spectrum. 
	The \textit{loer panel} shows the difference between the two spectra.}
        \label{compare}
\end{figure}

Pa$\alpha$ is clearly broad and certainly dominated by the Broad Line Region (BLR).
It is more difficult to indentify the presence of a narrow component associated
either with a Narrow Line Region (NLR) or with star formation.
A method to investigate the presence of a narrow component,
while exploiting the high angular resolution delivered by the AO,
is to study the spatial variation of the Pa$\alpha$ profile along
the slit. The BLR is certainly unresolved at our resolution (R$<$1pc) and
therefore the profile of the Pa$\alpha$, if totally attributed to the BLR,
should not change spatially along the slit, and should simply scale
in intensity following the profile of the PSF. We have investigated the
variation of the Pa$\alpha$ profile along the slit by
subtracting the continuum of all spectra at different locations
along the slit and then rescaled the spectra to match the Pa$\alpha$
flux on the central spectrum. In Fig.~\ref{compare} we show this comparison for
the spectrum $0.054''$ north of the central spectrum for PG1126: the
solid line shows the (continuum subtracted) central spectrum, while
the dashed line shows the rescaled off-nuclear spectrum. The lower
panel shows the difference between the two spectra which, although noisy,
shows the presence of a excess narrow component ($\rm FWHM \sim 200\ km/s$)
in the off-nuclear spectrum. This demonstrates the presence of
a narrow component of Pa$\alpha$,
associated either with a NLR or with star forming regions. 

\begin{figure}
	\centering
	\includegraphics[width=0.5\textwidth]{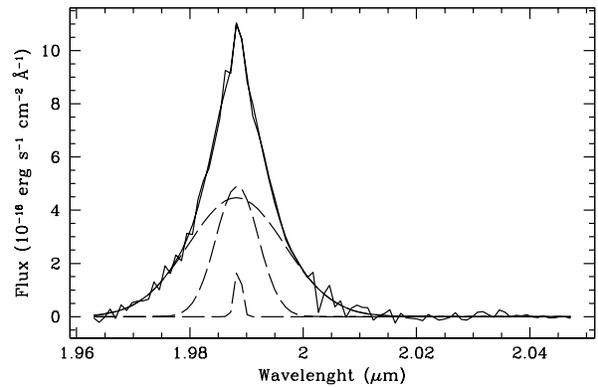}
	\caption{Fit of the Pa$\alpha$ profile for the central spectrum of PG112.
	The fit is overplotted 
	on the original spectrum, together with the single gaussian components (see text).}
        \label{fit}
\end{figure}

Once proved the existence of a narrow component, separated
from the BLR, we can follow its spatial behavior along with
other spectral features.
We have fitted the Pa$\alpha$ of PG1126 profile with two
broad gaussians, which reproduce the broad (BLR) component, and a narrow
component. The resulting fit for the central spectrum is shown in Fig.~\ref{fit}. 
\begin{table}
	\label{parametri}
	\begin{center}
	\begin{tabular}{l c c c c}
	\hline
	\hline
	Line & Flux & v$-$v$_{\rm sys}$ & FWHM  \\
	\hline
	\textbf{Pa$\alpha$} - Broad 1 & 953.3  $\pm$ 43.3   & 14 $\pm$ 37  & 3003 $\pm$ 134 \\
	\textbf{Pa$\alpha$} - Broad 2 & 457.1  $\pm$ 41.7   & 37 $\pm$ 32  & 1312 $\pm$ 142 \\
	\textbf{Pa$\alpha$} - Narrow & 27.6   $\pm$ 7.0    & 71 $\pm$ 32  & Unresolved \\ 
	\hline
	\textbf{Br$\delta$} - Broad  & 129.0  $\pm$ 10.4  & $-$75 $\pm$ 191 & 2277 $\pm$ 258 \\
	\textbf{Br$\delta$} - Narrow & 13.8   $\pm$ 6.3   & $-$124 $\pm$ 123  & 697  $\pm$ 413 \\
	\hline
	\textbf{[SiVI]} - Broad  & 89.7   $\pm$ 6.8   & $-$203 $\pm$ 333 & 2797 $\pm$ 240 \\
	\textbf{[SiVI]}	- Narrow & 8.0    $\pm$ 1.7   & 42 $\pm$ 89  & 360  $\pm$ 185 \\
  	\hline
	\hline
	\end{tabular}
	\caption{Best fit parameters for the components of the central spectrum of PG1126$-$041,
	for Pa$\alpha$, 
	Br$\delta$ and [SiVI]. \textit{Flux} is in units of $10^{-16}$ erg s$^{-1}$ cm$^{-2}$; 
	the shift the various component from the systemic velocity
	(\textit{v$-$v$_{sys}$}) and their \textit{FWHM} are in km s$^{-1}$.}
	\end{center}
\end{table}
In Tab.~2 we list the best fit parameters for the components. 
For what concerns the blend of [SiVI]1.96$\mu$m and Br$\delta$, each of the two lines
was fitted by using two gaussian components only (additional components are not
required given the lower S/N for these fainter lines)\footnote{Note that the profile
of the Pa$\alpha$ cannot be used to fit the Br$\delta$ because different hydrogen
lines originating from the BLR have often different profiles, like in this case.}. 
[SiVI] in particular is
well fitted by a broad and a narrow component. The presence of both narrow and broad
components of coronal lines (such as [SiVI]) is well known in AGNs. A component
with a width intermediate between BLR and NLR ($\rm \sim 1000-2000\ km~s^{-1}$)
is commonly observed and ascribed to an ``Intermediate Emission Line Region''
(Giannuzzo et al. \cite{giannuzzo95}). The narrow component of the coronal lines
comes instead
from the ``classical'' NLR and extends for a few 100~pc
(Maiolino et al. \cite{maiolino00}, Thompson et al. \cite{thompson01}).
The broad component requires also the profile to be asymmetric,
and specifically with a blue wing.
Such a blue asymmetry is also commonly observed
on high excitation lines from the nuclear region of several Seyferts
(e.g. Marconi et al. \cite{marconi96}, Oliva et al. \cite{oliva94}).

\begin{figure}
	\centering
	\resizebox{\hsize}{!}{\includegraphics{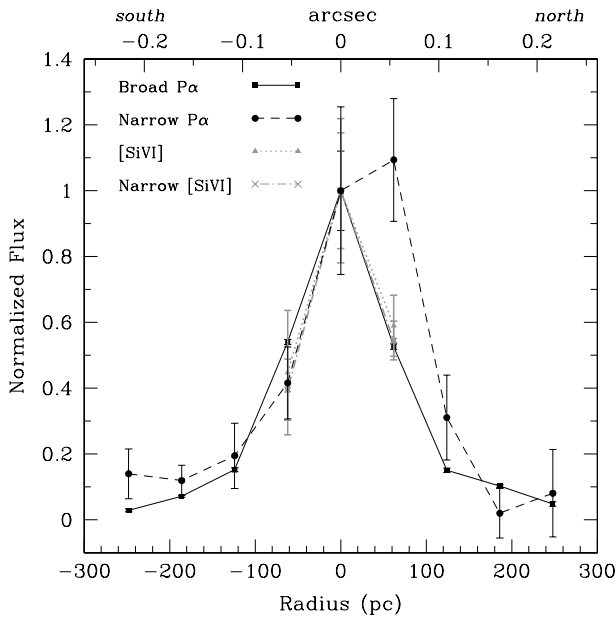}}
	\caption{Intensity of the various lines as a function of the offset from the nucleus, 
	normalized at the flux in the central spectrum. The peak emission of the narrow Pa$\alpha$ 
	is not aligned with the nucleus, but shows an excess $\sim$60 pc to the north.}
        \label{andamento}
\end{figure}

The line profiles on the off-nuclear spectra were fitted by keeping frozen
the parameters of the broad components (except for a scaling factor)
while the narrow components were left free to vary. In Fig.~\ref{andamento} we show
the intensity of the various lines as a function of the offset from the nucleus.  
The broad component of the Pa$\alpha$ gives the profiles of the
PSF along the slit (and it is also consistent with the profile of the
continuum, not shown). The most interesting result, expected from the
analysis discussed above, is that the intensity
of the narrow component of Pa$\alpha$ does not follow the PSF profile and shows an
excess of emission to the North, tracing an emission line region extending
for 0.05$''$--0.1$''$, or 60--120~pc, from the nucleus.

We investigate the nature of the narrow Pa$\alpha$ emitting region by 
comparing its flux with the observed [SiVI] emission. The narrow component
of the latter traces the NLR as it is not produced by star forming regions, 
while the narrow Pa$\alpha$ may either
come from star formation or from the NLR. The different trends of
the narrow Pa$\alpha$ and [SiVI] in Fig.~\ref{andamento} suggest that the two lines
have different origins. We have further investigated this issue by
comparing the ratio between narrow Pa$\alpha$ and [SiVI] in the
narrow line region of Sy2 galaxies observed by other authors
(Sy1 cannot be used because the hydrogen lines include the broad
component and generally a decomposition of narrow and broad components
is not given). Pa$\alpha$ is not observable
in low redshift Sy2s, but it can be easily derived from Br$\gamma$
by using the
case B recombination ratio (which holds for the NLR):
Br$\gamma$/Pa$\alpha$=0.082. The additional problem
is however that in Sy2 galaxies Br$\gamma$ may well be contributed by star
formation, since there are several evidences for nuclear and circumnuclear
star formation, as discussed in Sect.~1. For this reason we decided to focus
on two specific templates: NGC~1068 and Circinus. These are very
nearby Sy2's, extremely well studied, for which there is no evidence in the
central region for significant contribution to Br$\gamma$ by active star
formation (star formation occurs in rings at a radius of $\sim10''-30''$).
The Pa$\alpha$/[SiVI] ratio inferred for the NLR is 2.2 in Circinus
(Maiolino et al. \cite{maiolino98}) and 1.1 in NGC~1068
(Reunanen et al. \cite{reunanen03}).
In PG1126 the ratio between narrow Pa$\alpha$/[SiVI] is 3.4 on the
nucleus and 6.4 in the northern region at 60~pc from the nucleus. 
This comparison indicates that there is a strong excess of narrow Pa$\alpha$ 
emission with respect to that expected from a NLR.
Such a Pa$\alpha$ excess is probably 
due to active star formation in the nuclear and circumnuclear region, within
the central $\sim$100~pc.

Assuming that all the narrow Pa$\alpha$ is associated with star
formation we can estimate the star formation rate in the central
region of this quasar. Unfortunately, our data do not provide a
bi-dimensional information of the Pa$\alpha$ distribution
in the central region. However, we can roughly estimate the circumnuclear
integrated Pa$\alpha$ emission by assuming that the emission is well represented by the 
average of the northern and southern sides in our spectrum. The inferred total
narrow Pa$\alpha$ flux estimated in this way is
$\rm F_{tot}(Pa\alpha _{Narrow}) = 2.15~10^{-14}~erg~s^{-1}cm^{-2}$.
From case B recombination ($\rm Pa\alpha/H\alpha=0.107$), 
we infer a total H$\alpha$ luminosity due to star formation of
$\rm L(H\alpha)_{SF}=1.7~10^{42}~erg~s^{-1}$. By using the relation
between H$\alpha$ luminosity and star formation rate given in
Kennicutt et al. (\cite{kennicutt98}) we derive a nuclear star formation rate of
$\rm SFR \approx 13~M_{\odot}$ $\rm yr^{-1}$. According to the relation
between SFR and far-IR emission obtained by Kennicutt et al. (\cite{kennicutt98}),
such a star formation rate is expected to produce a far-IR
luminosity of $\rm \sim 3~10^{44}~erg~s^{-1}$, which matches the luminosity 
observed from this quasar (Tab.~1). Therefore, the nuclear star formation
detected in the nuclear few 100~pc may well account for most 
of the far-IR luminosity emitted by this quasar and could
explain the excess of far-IR emission in this object (Sect.~2). 
A similar result was found at higher redshift by Alexander et al. (\cite{alexander}),  
who derived that star formation appears to dominate the bolometric output 
of AGNs hosted in bright SCUBA galaxies.

Of course, the estimated nuclear star formation rate is derived from a
monodimensional spectrum and, in particular, the inferred nuclear (R$<$250~pc)
 Pa$\alpha$ luminosity has required a large aperture correction
(a factor of $\sim$3). As consequence, a confirmation of these findings is
certainly required with integral field spectroscopy. 
Another source of uncertainty is the possible star formation activity occurring in 
the host galaxy at radii larger than 250~pc; indeed the limited signal-to-noise 
ratio
of our spectra in the outer regions can only provide a relatively loose upper 
limit of 0.18 $\rm M_{\odot}~yr^{-1}~kpc^{-2}$ on the star formation rate per
unit surface area. Deeper observations are required to further constrain the
star formation in the host galaxy.

In the other two quasars the analysis of the line profile is more complex 
due to imperfect subtraction of some deep atmospheric absorption features 
and to a much lower Strehl ratio than in PG1126. 
There are some hints of a resolved narrow component of Pa$\alpha$, but need 
to be confirmed with higher quality spectra and higher Strehl Ratios.

\section{Conclusions}

We have obtained K-band spectroscopic observations assisted
by adaptive optics of three quasars at z$\sim$0.06
which are luminous in the far-IR. The adaptive optics correction
allows us the reach a diffraction limited angular resolution
($\sim0.08''$, corresponding to $\sim100$ pc),
with Strehl ratios ranging from 5\% to 26\%.
In the quasar observed with the highest Strehl ratio (PG1126$-$041)
we spatially resolve a narrow component of Pa$\alpha$ on scales
of $\sim$50--100~pc. By comparing the spatial distribution of the
narrow Pa$\alpha$ with the distribution of the coronal line
[SiVI]1.96$\mu$m, we infer that most of the narrow Pa$\alpha$ is
due to star formation in the nuclear and circumnuclear region
of the quasar. We also derive that most of the far-IR emission
of this quasar is due to the nuclear star formation detected by us.

These observations clearly demonstrate that adaptive optics assisted
spectroscopy is a powerful tool to investigate the nuclear region
even in bright quasars.

\begin{acknowledgements}
      This work was partially supported by
      the Italian Ministry of Research (MIUR).
      We are grateful to the ESO staff on Paranal for having
      performed these observations in service mode.
\end{acknowledgements}

\end{document}